# Metallic Diluted Dimerization in $VO_2$ Tweeds

*Felip Sandiumenge*, Laura Rodríguez, Miguel Pruneda, César Magén, José Santiso, Gustau Catalan*

Dr. F. Sandiumenge
Institut de Ciència de Materials de Barcelona (ICMAB-CSIC)
Campus de la UAB
08193 Bellaterra, Catalonia, Spain
E-mail: felip@icmab.cat

L. Rodríguez, M. Pruneda, J. Santiso, G. Catalan
CN2 (Institut Catala de Nanociencia i Nanotecnologia) BIST-CSIC
08193 Bellaterra, Catalonia, Spain

C. Magén
Instituto de Ciencia de Materiales de Aragón (ICMA)
Universidad de Zaragoza - CSIC
Departamento de Física de la Materia Condensada
50009 Zaragoza, Spain
and
Laboratorio de Microscopías Avanzadas (LMA) - Instituto de Nanociencia de Aragón (INA), Universidad de Zaragoza
50018 Zaragoza, Spain

G. Catalan
ICREA (Institució Catalana de Recerca i Estudis Avançats)
08010 Barcelona, Catalonia

The observation of electronic phase separation textures in vanadium dioxide ($VO_2$), a prototypical electron-correlated oxide, has recently added new perspectives on the long standing debate about its metal-insulator transition and its applications. Yet, the lack of atomically-resolved information on phases accompanying such complex patterns still hinders a comprehensive understanding of the transition and its implementation in practical devices. In this work, atomic resolution imaging and spectroscopy unveils the existence of ferroelastic tweed structures on ~5 nm length scales, well below the resolution limit of currently employed spectroscopic imaging techniques. Moreover, density functional theory calculations show that this pre-transitional fine-scale tweed, which on average looks and behaves like the standard metallic rutile phase, is in fact weaved by semi-dimerized chains of vanadium in a new monoclinic phase that represents a structural bridge to the monoclinic insulating ground state. These observations provide a multiscale perspective for the interpretation of existing data, whereby phase coexistence and structural intermixing can occur all the way down to the atomic scale.

Symmetry-breaking at metal-insulator transitions (MITs) typically results in the formation of spatial patterns which enclose information on the nature of such transformations.[1,2] The paradigm of such transitions is vanadium dioxide ($VO_2$), a $3d^1$ system that at ~ 340K undergoes a first-order MIT between a tetragonal (rutile) metal and a monoclinic insulating phase,[3] which makes this compound attractive for a number of device applications.[4,5] The transition is mediated by dimerization of V ions along the V-chains of the rutile structure, and despite the relative contributions of electron-electron correlations, electron-lattice instabilities or lattice vibrations to the opening of the ~0.6 eV band gap have long been a focus of interest,[6-9] the strong sensitivity of $VO_2$ to external stimuli and the occurrence of transient competing states still makes its understanding and implementation of the above mentioned applications difficult.

The observation of weakened Coulomb correlations preceding the structural change,[10] in tune with numerous experimental and theoretical evidences for the coexistence of V-V dimers with metallicity in monoclinic states, has led to numerous speculations about the Peierls/Mott character of the transition.[11-21] A key aspect on the understanding of the MIT is the way this decoupling manifests in real-space patterns. Observations of metal/insulator phase coexistence reported so far can be classified into two types: i) spectroscopic imaging revealing transient nanoscale electronic phase separation within low-symmetry (monoclinic) phases in strained films,[17-20] and ii) martensitic-like (mesoscopic) domain patterns in freestanding beams as well as in epitaxial films,[19,22-24] combining phases related by the thermodynamic *P*-*T* triple point.[9,25-27] Taken together, these observations point to a multiscale pattern evolution across the transition that has remained elusive to date.

Coexisting nano- and mesoscale patterns can have a common origin in pre-transitional spatial fluctuations such as those leading to tweed or ferroic-glass textures in ferroelectrics,[28,29] high-temperature superconductors [30,31] or metallic alloys.[32,33] A similar scenario already successfully explains multiscale, multiphase texturing in perovskite manganites.[2] It is plausible that, below the 20-40 nm resolution range accessed by spectroscopic imaging,[18,20] the coupling of ferroelasticity with misfit strain can drive the formation of relaxation patterns on even finer length scales in $VO_2$. Indeed, combining high resolution transmission electron microscopy (HRTEM), electron energy loss spectroscopy (EELS) and density functional theory (DFT) calculations, here we uncover the existence of $VO_2$ tweed textures with characteristic correlation lengths of ~5 nm, and disentangle its local symmetry.

**Figure 1a** captures the space-average structure of the $VO_2/TiO_2(001)$ films studied in this work (see Supplemental Information, Figure S1). The presence of $(4\ 0\ \text{-}2)_{M1}$ and $(0\ 0\ 2)_R$ reflections in the θ-2θ scans of the 40 nm and 80 nm -thick films suggests the coexistence of the M1 and rutile phases. At a thickness of 90 nm, the unique presence of the M1 peak indicates that the whole volume has evolved to the equilibrium monoclinic phase. Relaxation of the in-plane tensile stress imposed by the substrates proceeds through cracking rather than misfit dislocation,[34-36] and Figure 1b illustrates a typical stress-relieving crack pattern in a 60 nm -thick film. The metal/insulator character of the participating phases can be discerned in reflection optical microscopy images according to their different optical density.[22,34] Thus, the bright contrast (lower optical density) decorating the cracks corresponds to the insulating

M1 phase, while the dark (higher optical density) rectangular tile-like regions between the cracks correspond to metallic $VO_2$.[34] This behavior is further confirmed by conducting-AFM current maps and I(V) measurements, the latter indicating ohmic (metallic) conduction even below room temperature (see Supplemental Information, Figure S2). Reciprocal space mapping around the $(-1\ -1\ 2)_R$ reflection indicates that in the 40 and 80 nm -thick films both the R and M1 phases are fully in-plane strained (Figures 1c,d), while in the 90 nm -thick film the M1 phase is in-plane relaxed (Figure 1e). The HRTEM analysis presented below was performed in regions far from the cracks in the 40 nm and 80 nm -thick films; these are the regions that appear optically metallic and crystallographically rutile-like according to X-ray diffraction. An HRTEM analysis on the 90 nm -thick film is presented as Supplementary Information (Figure S3).

Figure 2a shows a cross-section HRTEM image of the 40 nm-thick $VO_2$ film viewed along the $[100]_R$ direction. On a fine scale, we notice segments of bright fringes parallel to two prominent directions normal to the arrows labeled x3a and x3b. In the enlarged image depicted in Figure 2b it can be clearly seen that the fringes triple the periodicity of the dot pattern along the x3a and x3b directions, without any appreciable alteration in the positional order of the atomic columns, which still define a body centered tetragonal lattice consistent with the rutile structure (yellow rectangle). Such fringes are grouped in nanodomains 3-6 nm in size. In agreement with these features, the brightest spots in the fast Fourier transform (FFT) can be assigned to the $VO_2$ rutile structure oriented along the $[1\ 0\ 0]_R$ zone axis (Figure 2c). All other weaker spots (yellow circles) correspond to the tripled periodicity of the $(0\ 1\ 1)_R$ and $(0\ -1\ 1)_R$ rutile planes, and can be respectively assigned to the orientational variants denoted by x3a and x3b. Domains with orientations x3a and x3b are related by a mirror plane, $m_y$, parallel to $(0\ 1\ 0)_R$ of the underlying rutile ordering (Figures 2b,c). This symmetry operation becomes, therefore, implicitly excluded from the ordered x3 phase and reduces its symmetry to either monoclinic or triclinic. Figure 2d shows a filtered image of the same area obtained by selecting $(0\ \frac{1}{3}\ \frac{1}{3})_R$ and $(0\ \frac{2}{3}\ \frac{2}{3})_R$ spots. The contrast of this image thus maps the spatial distribution of the x3 ordering and defines a tweed texture in which interwoven fringes are attributed to the overlapping of the two orientations along the viewing direction.

We can thus infer that, during cooling, the in-plane strained rutile phase undergoes a tweedy texturing transition driven by the symmetry-breaking induced by the x3 order. Since this process preserves the average structure of the precursor rutile template, it is not expected to induce noticeable changes in the X-ray diffraction data presented in Figure 1. Thus, while X-ray diffraction shows this phase to be rutile, and optical microscopy and I(V) curves also show it to be metallic, TEM shows that this is NOT the standard high-temperature metallic rutile phase of $VO_2$, but a metallic tweed with a distinct x3 local order.

At this point it is pertinent to comment on the sense of the *tweed* term. According to its most classical definition, *tweed* refers to spatial fluctuations of the order parameter, leading to densely interwoven embryos of the low symmetry phase.[37] In contrast, in the present case the origin of the tweed texture lies on the formation of an intermediate phase which is different than the low symmetry M1 ground state. Thus, as will be shown below, the tweed domains exhibit the fully developed order of this intermediate phase. Parenthetically, a similar mechanism has been reported in martensitic Au-Cu-Al alloys.[38]

As a first step to investigate the possible origins of the x3 order, we acquired atom-column resolved EEL spectra at the V-$L_{2,3}$ edge along the rutile $c_R$-axis direction. This absorption edge probes transitions from the V 2p core level to the 3d valence electronic states, $2p^6 3d^1 \rightarrow 2p^5 3d^2$, and provides a means for the determination of charge variations at high spatial resolution by examining the position and intensity ratio of the V-$L_{2,3}$ white lines.[6] We didn't detect variations in peak positions above ~0.1 eV. Taking into account that the addition of one electron to the V 3d levels carries a spectral shift of ~1 eV,[39] we conclude that the x3 order is not associated to variations larger than 10% in the electric charge of V ions, thereby excluding strong $V^{4+}/V^{3+}$ charge or associated vacancy ordering effects. On the other hand, it is well known that ordered oxygen vacancies in $VO_2$ cause Magnéli phases.[40] We calculated the electron diffraction patterns along orientations compatible with the orientation of our films for the different Magnéli phases and did not find any matching with the experimental FFT.

On a structural basis, the problem of tripling the periodicity of the $(011)_R$ planes reduces to calculating the number of ways three bonds, two of which having the same length, can be combined along the rutile $c_R$ axis (Figures 2e-h). At the rutile→M1 transition, the V chains become fully dimerized as can be seen by comparing Figures 2e,f. The two triple models can be derived from the M1 structure by a simple rearrangement of the long (L) and short (S) V-V distances, leading either to a diluted dimer structure (Figure 2g) or to a trimer configuration (Figure 2h). We note, however, that the trimerization of V ions hypothesized in Figure 2h is unlikely as V-trimers have only been predicted in triangular configurations associated to Frenkel defects.[41] Our own calculations, discussed later in the article, also argue against this scenario.

In order to get further insight into the electronic structure of the x3 ordered phase, Figure 3a compares the EELS V-$L_{2,3}$ and O-K edges obtained from the tweed pattern and the M1 phase (90 nm -thick film, see Figures 1a,e and Supplementary Figure S3). The V-$L_{2,3}$ edges look essentially similar, indicating that there are no significant differences between the oxidation state of V ions in both samples. In particular, oxygen vacancies have been shown to manifest as a lowering in the intensity of the $L_3$ and O-K $\sigma^*$ signals.[42] Therefore, the marked similarity between the $L_3/L_2$ intensity ratios in the ordered and M1 phases, and the leveling of the O-K $\pi^*$ and $\sigma^*$ intensities in the former (see below), altogether further confirm that the observed tweed is not associated with any significant amount of oxygen vacancies. These edges are, in fact, dominated by the strong interaction between the 2p core-hole and the 3d electrons in the excited state and are less sensitive to changes in the density of states than the O-Kedge, which probes transitions from the O 1s core level to unoccupied 2p states hybridized with V 3d levels.[6,43]

In the fully dimerized M1 structure, the O-K edge is composed of three features, as indicated in Figure 3a: π*, $d_{\parallel}$*and σ*.[44] Looking at Figure 3a it can be clearly seen that the main difference between the M1 and x3 spectra is a leveling of the intensities of the π* and σ* peaks. Since EELS provides information on the density of unoccupied states above the Fermi level, this decrease of the π* intensity indicates a higher occupancy of this band, which, as we discuss below, agrees with the metallic behavior observed by optical microscopy (Figure 1b). On the basis of DFT calculations using fixed long to short V-V distance ratios, a similar but more pronounced relative supression of the π* intensity in strained 10nm thick films has been attributed to a homogenization of V-V distances within a monoclinic structure.[45] Below we show that our results are consistent with an ordered dilution of V-V dimers also leading to metallicity in monoclinic $VO_2$.

In order to further inquiry into the atomic and electronic structures associated to the x3 ordering we performed DFT-based atomistic calculations. Although no general consensus exists on the reliability of band-theory methods to predict energy gaps and magnetic orderings of $VO_2$ phases,[46-48] including an electronic Hubbard repulsion term on vanadium 3d orbitals has been shown to give accurate results for structural and vibrational properties.[9,49] Here, we use a $U$=4 eV on top of the PBE exchange-correlation functional. In agreement with previous results, in the harmonic approximation we observed the presence of soft phonons in the rutile phase that have been used to explain the transition to the M1 phase.[49] A finite-differences phonon calculation using a supercell compatible with the x3 periodicity gives a soft phonon mode that favors dimerization with an intermediate “isolated” V site. Even though these soft-phonons could be stabilized by anharmonic contributions,[9] guided by the above experimental findings, we performed full relaxation of the structure following the vibration eigenmodes, resulting in a metallic phase with two out of three V ions forming dimers along $c_R$-axis. The V-V intra-dimer distance is ~2.51 Å, only slightly longer than the dimer bond length obtained by computation of the relaxed M1 structure, ~2.50 Å, which is in good agreement with previously reported theoretical results based on similar methods.[9] The “isolated” V site is 2.9-3.0 Å apart from the dimerized V ions. The dimer ordering of this phase thus differs from those found in the insulating monoclinic M1, M2 and triclinic (T) polymorphs,[50-52] in which at least one V-chain sublattice is fully dimerized. Energetically and crystallographicallly, then, this metastable structure (hereafter referred to as x3M) falls between the rutile and M1 phases and fits a transitional bridging role.

Only a tiny modulation of the electronic charges can be observed in the Mulliken atomic populations (less than 1%). The projected density of states on the V 3d states (Figure 3b) shows that dimerization, as for the M1 phase (Figure 3c), induces the splitting of the $d_{\parallel}$ band into bonding and antibonding (unoccupied) peaks, but the splitting is dramatically suppressed in the “isolated” V site. In contrast with previously proposed monoclinic metallic structures, in which the metallicity results from homogeneous bonding distortions relative to the M1 structure (suppression or distortion of the V-V dimer tilts[15,16] or the homogenization of short and long V-V distances[13,21,45]), in the x3M structure the essential effect, despite unequal distortions in the two chain sublattices, is the dilution of V-V dimers by alternating "isolated" V ions (Figures 3b,c, and Supplementary Table S4 and Figure S4 for crystal structure data). In this situation, the depopulation of the $d_{\parallel}$ state in the latter site translates into an increased occupancy of the $\pi^*$ state consistent with the relative decrease of the $\pi^*$ intensity observed in the EEL spectrum (Figure 3a). In other words, the calculated partial dimerization, resulting in the observed triple-periodicity, is not sufficient to open an insulating gap.

Image simulations (defocus 91nm, thickness 10nm) of the model derived from DFT calculations show excellent agreement with the HRTEM experimental image (Figure 3d). Assigning the 2/*m* point symmetry to the x3M structure, the symmetry lowering relative to the rutile one would generate four equivalent domains making the $[1\ 0\ 3]_{x3M}$ and $[0\ 1\ 0]_{x3M}$ orientations equally probable along the $[1\ 0\ 0]_R$ zone axis, in agreement with the image simulations labeled ❶ and ❷ shown in Figure (see Supplementary Figure S5a).

As a consequence, it is the inter-weaving of these four energetically-equivalent orientations that results in the HRTEM contrast details of the tweed observed in Figure 2a (see also Supplementary Figure S6a). Figures 3e,f show that the corresponding theoretical electron diffraction patterns (EDPs) also show excellent agreement with the FFT shown in Figure 2c (see also Supplementary Figures S5a,b).

Figure 4a shows a HRTEM image of the 80 nm -thick film. The corresponding FFT pattern (Figure 4b) is similar to that shown in Figure 2c, but incorporates a set of additional spots at the $(0\ ½\ ½)_R$, $(0\ 1\ 0)_R$ and $(0\ 0\ 1)_R$ positions. Furthermore, the x3M spots are more diffuse and the background level is higher, indicating a higher structural disorder than in the 40 nm -thick film. The occurrence of the $(0\ ½\ ½)_R$, $(0\ 1\ 0)_R$ and $(0\ 0\ 1)_R$ spots is consistent with the unit cell doubling associated to the M1 and M2 polymorphs, both connected by instabilities at the same $(0\ ½\ ½)_R$ point of the rutile Brillouin zone[53] (Supplementary Figures S4b,c). The double and triple rutile $(0\ 1\ 1)_R$ periodicities observed in the filtered image depicted in Figure 4c demonstrate the coexistence of [0 1 0] oriented domains of both the M1 and x3M phases. Regions where the contrast modulations vanish are difficult to identify owing to the similitude between some projections of the coexisting structures (see Supplementary Figure S6). This is evident by observing the strong similitude between the calculated EDPs of the M1 and M2 domain orientations along the rutile $[1\ 0\ 0]_R$ zone axis (Figs. 4d-g). The additional spots referred to above are marked in red and indexed in their respective settings in the magnified view of the FFT shown in the right panel of Figure 4b).

According to these observations, the sequence of microstructures observed in the 40 and 80 nm -thick films can be explained by the transformation of the rutile structure into tetragonal-like templates (aka “metrically tetragonal”), formed by coincident-site lattices of nanoscale ferroelastic domains of the three monoclinic phases. In the thinner 40 nm film the dominant phase is a transient monoclinic x3M metallic tweed, while the complex structure of the 80 nm thicker film can be described as a cluster composite of x3M, M1 and M2 phases coexisting on a ~5 nm length scale. Therefore, the electronic phase coexistence in $VO_2$ occurs at length scales well below those accessed by currently employed X-ray spectroscopic imaging techniques, which, combined with the prevalence of the rutile-like structural scaffolding, may explain why this complex structural behavior had been missed.

The scenario drawn above, reminiscent of the evolution from tweed to glassy textures observed in a number of ferroelectric, ferroelastic and martensitic systems,[28-33] is similar to that suggested to explain multiphase texturing in perovskite manganites.[2] Our calculations show that it is the suppression of the bonding/antibonding splitting in the isolated V site bridging consecutive V-V dimers, and the subsequent depopulation of this level, what governs the metallicity in the monoclinic x3M structure. We note that, in contrast with other studies, the x3M structure has been obtained in the absence of symmetry constrains, which makes sense in the present highly disordered tweeds. The multicomponent tweeds uncovered in this work show that previously reported monoclinic metal/insulator phase coexistence in $VO_2$ probably corresponds to pretransitional stages of its ferroelastic-martensitic transformation pointing to a strong participation of lattice degrees of freedom.

## Experimental Section

*Growth of films*: $VO_2$ films with thicknesses of ~40 nm, ~80 nm and ~90 nm were grown by PLD, using a $V_2O_5$ ceramic target (99.9% purity) provided by Kurt J. Lesker. The ablation was performed by a KrF excimer laser, Lambda Physik COMPex 201, emitting at a 248 nm wavelength. The optimal growth conditions were: 400 °C deposition temperature, 10 mTorr gas pressure of pure $O_2$, laser fluence of 0.8 $J \cdot cm^{-2}$ and a working distance of 55 mm from target to substrate. The cooling down step was performed at 15 $°C \cdot min^{-1}$ keeping the same gas pressure. The thickness of the films was controlled by adjusting the number of pulses (at these conditions the growth rate was ~0.03 nm·pulse$^{-1}$). The single crystal substrates were rutile $TiO_2$ (001) (and metallic Nb-doped $TiO_2$ for local transport measurements) provided by CrysTec. This substrate induces an in-plane biaxial tensile strain both in the rutile and monoclinic $VO_2$ phases. In the rutile-$VO_2$ case, the misfit strain is ~+0.9%. In the M1-$VO_2$ case, the misfit strain is ~+1.7% and ~+1.4% along the [0 1 0] and [1 0 2] in-plane directions, respectively.

*X-ray diffraction (XRD)*: θ-2θ scans scans and reciprocal space maps were measured in a Lab diffractometer (Malvern-Panalytical X'pert Pro MRD) equipped with a 2x Ge(110) monochromator (Cu $K\alpha_1$ radiation, λ= 1.5406 Å) and a four-angle goniometer.

*Conducting atomic force microscopy*: Local electrical conductivity maps and I(V) curves were measured at 15°C by conducting-AFM (Asylum MFP-3D) using a Pt/Ir tip provided by Nanosensors. Films for c-AFM measurements were grown on conducting Nb-doped $TiO_2$ substrates. Measurements were performed in a perpendicular current configuration using a sample voltage of -23 mV and the scans show the amount of current that passes from the bottom to the top of the film as detected by the grounded AFM tip.

*Transmission electron microscopy*: Cross-sectional TEM lamellae were prepared by $Ga^+$ Focused Ion Beam procedures. High-Resolution Transmission Electron Microscopy (HRTEM) images were acquired at 300 kV using Thermo Fisher Scientific Titan Cube 60-300 and Hitachi HF3300 field emission gun microscopes, equipped with CEOS imaging aberration correctors. Both instruments provide a spatial resolution < 1 Å. Electron Energy Loss Spectroscopy (EELS) was performed in Scanning Transmission Electron Microscopy mode using a Thermo Fisher Scientific Titan 60-300 microscope equipped with a high-brightness Schottky field emission gun (X-FEG), a Wien filter monochromator and an aberration corrector for the condenser system to achieve a probe size < 1 Å in STEM mode. We employed a Tridiem 866 ERS image filter/spectrometer from Gatan. Cumulative EELS spectra were collected while fast-scanning a small area of the sample (approximately 2x2 $nm^2$) to minimize beam damage. An energy dispersion of 0.1 eV and a collection angle of 43 mrad were selected. Single spectrum acquisition time was 1 s, while the accumulated exposure time ranged from 50 to 100 s. Multislice image simulations were performed using the jems software package.[54]

*Density functional theory calculations*: Calculations were carried out using a numerical atomic orbitals approach implemented in the SIESTA code.[55,56] We used the Perdew–Burke–Ernzerhof functional[57] to account for the exchange-correlation energy, and included an effective Hubbard term $U$ = 4eV in V 3d atomic orbitals, which gives a band gap in the M1 phase which is in agreement with previous theoretical reports. A split-valence double-ζ basis set was used to describe the valence electrons wave function, whereas the core electrons were replaced by norm conserving scalar relativistic pseudopotentials factorized in the Kleinman–Bylander form.[58] We used an energy cutoff of 1200 Ry for the real space integration, and a Monkhorst–Pack $k$-point grid of 9 × 9 × 9 to account for the sampling of the Brillouin zone. Tolerances of $10^{-6}$ and $10^{-4}$ on the density matrix and total energy, respectively, were used for the convergence of the self-consistency. Geometrical optimizations were performed for fully relaxed cells, and imposing strains compatible with experimental conditions (films grown on $TiO_2$-(001) under tensile strain), until atomic forces were smaller than $1 \times 10^{-2}$eV/Å$^{-1}$.

**Acknowledgements**

We thank Prof. Enric Canadell for enlightening discussions. We also thank Prof. Etienne Snoeck and Dr. Christophe Gatel for obtaining HRTEM images of the 80 nm-thick film at CEMES (Toulouse), and Dr. Belén Ballesteros for assistance during preliminary TEM observations. F.S. acknowledges the Spanish Ministry of Industry, Economy and Competitiveness (MINECO) for financial support through the “Severo Ochoa” programme for Centers of Excellence in R&D (SEV-2015-0496) and project no. RTI2018-098537-B-C21. G.C., J.S. and L.R. acknowledge financial support from MINECO (grant MAT2016-77100-C2-1-P) and the Catalan Government (Generalitat de Catalunya, grant 2017 SGR 579 ). M.P. acknowledges financial support from Spanish MICIU, AEI and EU FEDER (Grant No. PGC2018-096955-B-C43) and Generalitat de Catalunya (Grant No. 2017SGR1506 ). Research at ICN2 is also supported by the CERCA programme (Generalitat de Catalunya) and by the "Severo Ochoa" programme for Centers of Excellence in R&D of MINECO (grant SEV-2017-0706). C.M. acknowledges funding from MINECO under project MAT2017-82970-C2-R, and from the European's Union Horizon 2020 research and innovation programme under Grant No. 823717-ESTEEM3.

Received: ((will be filled in by the editorial staff))
Revised: ((will be filled in by the editorial staff))
Published online: ((will be filled in by the editorial staff))

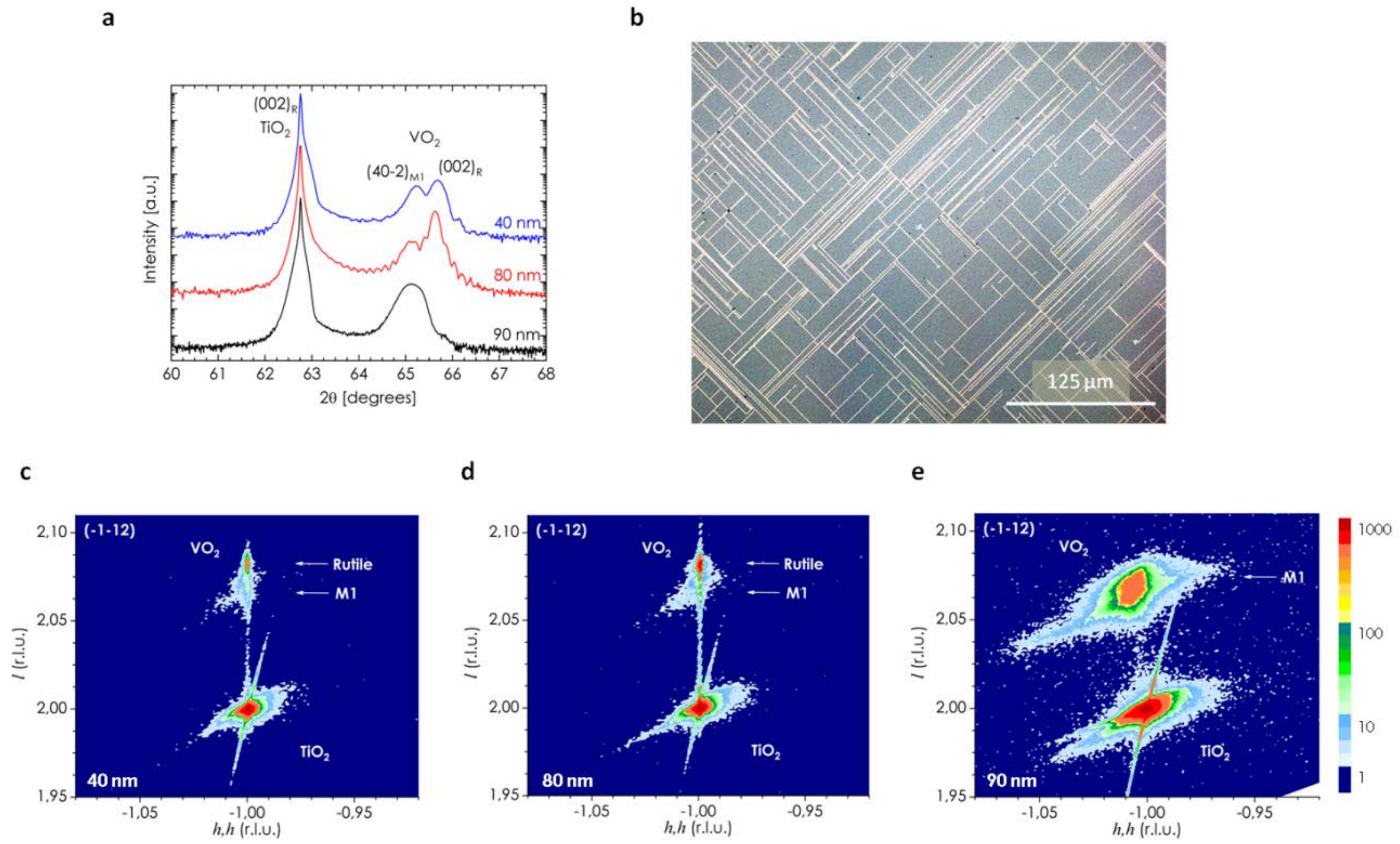

**Figure 1.** (a) θ-2θ scans of three $VO_2$ films with thicknesses of 40 nm, 80 nm and 90 nm. The presence of $(4\ 0\ {-2})_{M1}$ and $(0\ 0\ 2)_R$ reflections suggest the coexistence of the monoclinic M1 and tetragonal rutile-like phases in the 40 nm and 80 nm -thick films, while the 90 nm -thick one only exhibits the monoclinic $(4\ 0\ {-2})_{M1}$ peak. (b) Reflected-light optical micrograph of the surface of a 60 nm -thick film showing a typical crack framework along the rutile <110> directions . Bright stripes decorating the cracks correspond to the insulating M1 phase of the relaxed lattice. By contrast, the tile-like rectangular regions bounded by these cracks, are metallic.[34] (c-e) Reciprocal space maps obtained around the substrate $({-1}\ {-1}\ 2)_R$ reflection. For the 40 nm (c) and 80 nm (d) -thick films, the in-plane film components of both the M1 and rutile-like reflections are vertically aligned with the substrate one. Conversely, in the 90 nm -thick film, the M1 and substrate in-plane components are vertically misaligned indicating a relaxed state.

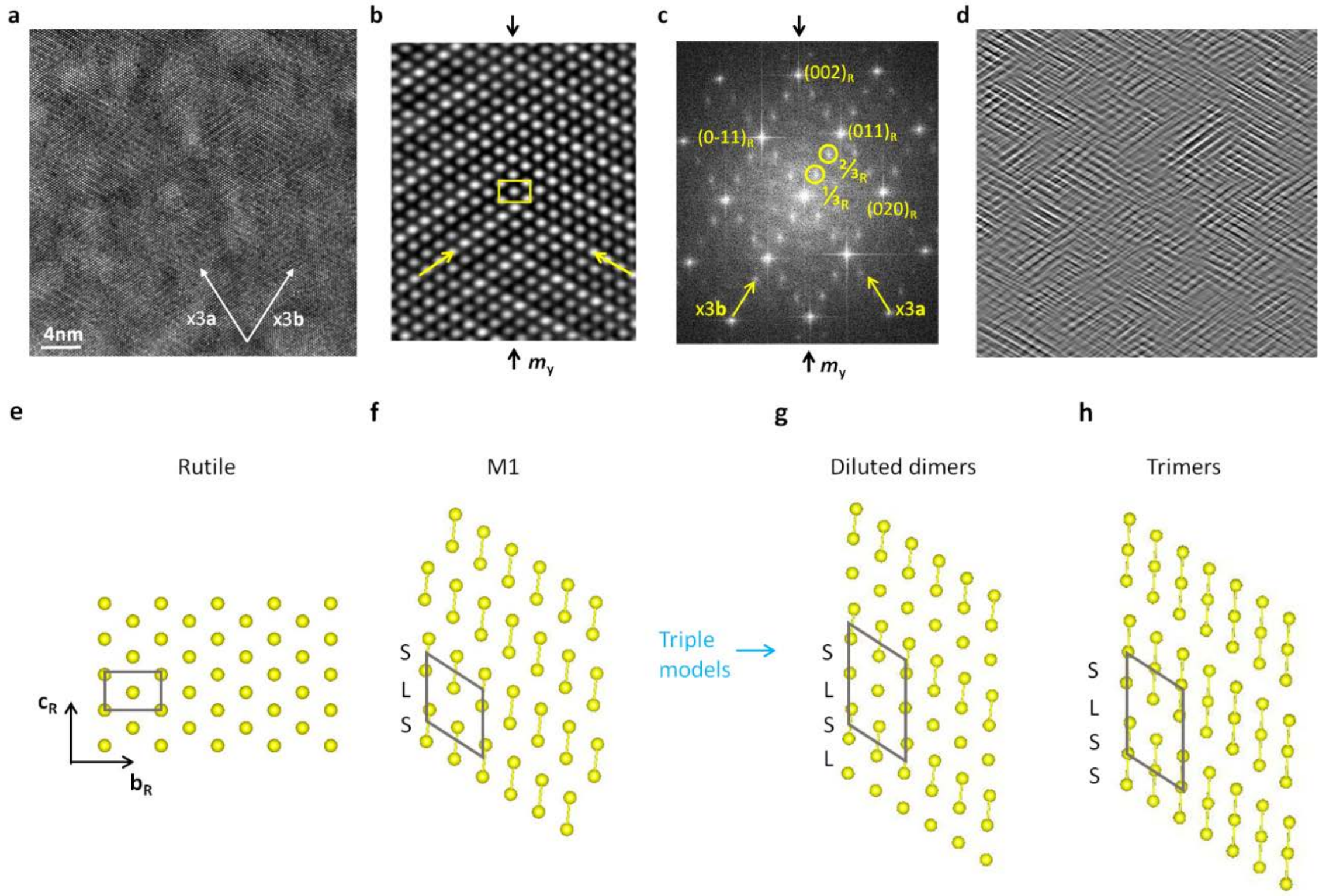

**Figure 2.** (a) HRTEM raw image of the tweed texture observed in the 40 nm -thick film viewed along the rutile $[100]_R$ zone axis. The contrast modulation directions are indicated by arrows labeled x3a and x3b. (b) Filtered enlarged view of (a). Arrows indicate the brightest atomic rows defining the contrast modulation. The yellow rectangle highlights the rutile-like body-centered tetragonal arrangement of atomic columns underlying the contrast modulation. The two modulation directions are related by a $(0\ 1\ 0)_R$-type mirror plane, $m_y$, of the rutile-like lattice. (c) FFT pattern obtained from the image shown in (a). The pattern is indexed with reference to the rutile lattice. Superestructure spots arising from the contrast modulation observed in (a) and (b), $⅓_R$ and $⅔_R$, correspond to $(0\ ⅓\ ⅓)_R$ and $(0\ ⅔\ ⅔)_R$, respectively. The $m_y$ mirror plane is indicated by vertical arrows. Yellow arrows correspond to the modulation directions indicated in (a,b). (d) The same image shown in (a), filtered using the $⅓_R$ and $⅔_R$ superestructure spots to highlight the spatial distribution of ordered domains. (e,f) Schematic representations of the V-ion arrangement in the rutile and monoclinic M1 lattices projected along the $[1\ 0\ 0]_R$ direction. (g,h) Schematic representations of hypothetic V-V dimer arrangements along the $c_R$-axis direction satisfying the $⅓_R$ and $⅔_R$ superestructure spots

observed in the FFT. Long and short V-V distances along the $c_R$-axis directions are indicated in (b-h) by "L" and "S", respectively. Unit cells are indicated with black lines in (e-h).

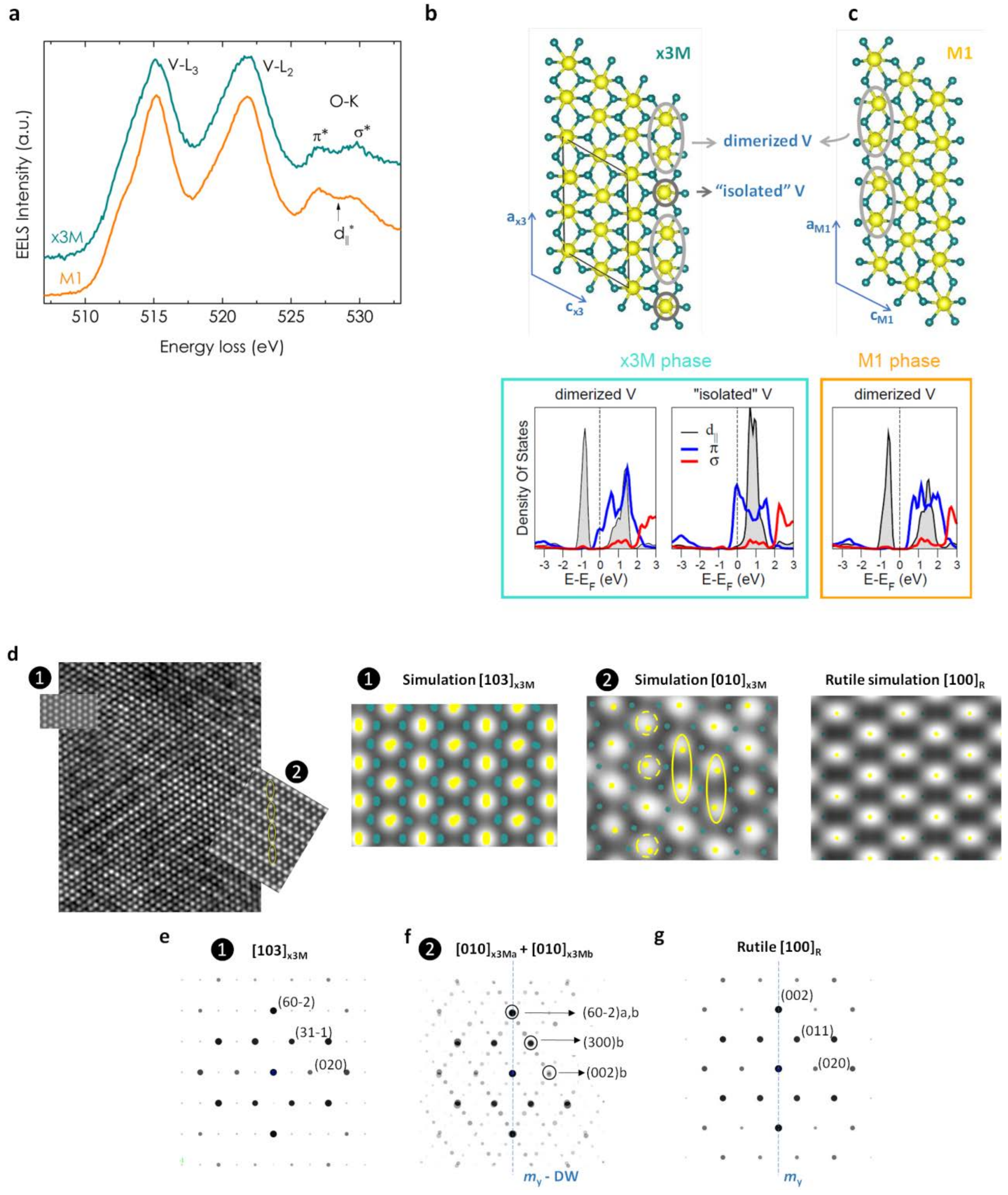

**Figure 3.** (a) V-$L_{2,3}$ and O-K edge EEL spectra obtained from the tweed texture shown in Figure 2a, representative of the x3 phase (blue dots), and from the fully relaxed 90 nm -thick film, representative of the M1 phase (red dots). The dimerization of V chains splits the $d_{\parallel}$ band into a fully occupied bonding ($d_{\parallel}$) and an empty anti-bonding ($d_{\parallel}$*) combination, and the latter appears about 0.7 eV above the π* band edge.[6] The lack of a resolved $d_{\parallel}$* signal in the EEL

spectrum of the M1 sample is a consequence of the overlapping with the more isotropic π* and σ* orbital structures due to the isotropic scattering cross-section inherent to the EEL experiment. (b,c) Projected Density of States on the V 3d states for vanadium atoms in the “dimerized” and “isolated” sites for the x3M (b) and M1 (c) structures. The $d_{\parallel}$ states (shaded grey curve) form bonding/antibonding states for dimerized V ions, but the splitting is suppressed for the "isolated V" ion, resulting in partial occupation of the π* states (blue curve) in the x3M phase. (d) Left panel: Image simulations of the x3M structure obtained from DFT calculations overlaid on the row experimental HRTEM image, for projections along the $[1\ 0\ 3]_{x3M}$, ❶, and $[0\ 1\ 0]_{x3M}$ ❷, zone axes. The two panels are enlarged views of the simulations showing atom positions (V and O atoms are shown yellow and green, respectively). In the $[1\ 0\ 3]_{x3M}$ projection, the overlapping of V atoms belonging to adjacent V-chains produces an image similar to that of the rutile structure projected along the $[1\ 0\ 0]_R$ (or $[0\ 1\ 0]_R$) zone axes. In the simulation corresponding to the $[0\ 1\ 0]_{x3M}$ zone axis, V-V dimers are indicated by yellow ellipses. As highlighted by yellow broken circles, within the dimers the V atoms are offset towards their center, resulting in apparently larger V-V distances. The right panel shows a simulation of the rutile structure along to the $[1\ 0\ 0]_R$ (or $[0\ 1\ 0]_R$) zone axis. (e) $[1\ 0\ 3]_{x3M}$ zone axis calculated electron diffraction pattern of the x3M phase corresponding to the image simulation labeled ❶. (f) Superposition of electron diffraction patterns corresponding to domains x3Ma and x3Mb, related by the missing rutile $m_y$ mirror plane. These orientations correspond to the image simulation labeled ❷. The patterns shown in (e) and (f) are indexed with reference to the x3M structure. (g) $[1\ 0\ 0]_R$ (or $[0\ 1\ 0]_R$) zone axis electron diffraction pattern of the rutile structure.

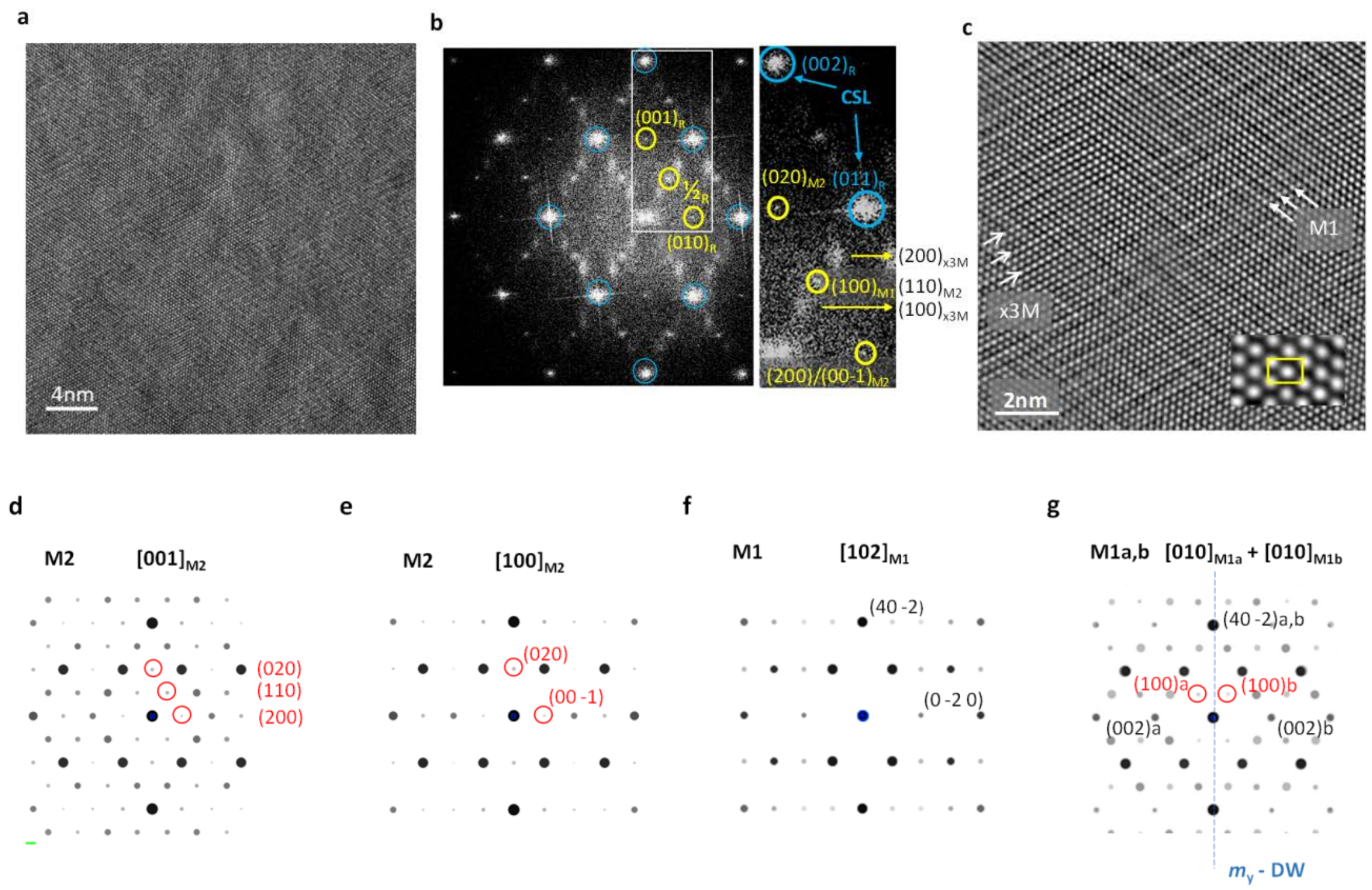

**Figure 4.** (a) HRTEM raw image of the tweed texture observed in the 80 nm -thick film viewed along the rutile $[100]_R$ zone axis. Contrast modulations are weaker and exhibit smaller correlation lengths than in images obtained from the 40 nm -thick film. (b) FFT pattern obtained from the image shown in (a). Yellow circles mark additional spots: $(0\ 1\ 0)_R$, $(0\ 0\ 1)_R$, $(0\ ½\ ½)_R$ (labeled $½_R$) with respect to the FFT obtained from the 40 nm -thick film (Figure 2c). The right panel is an enlarged image of the region enclosed within the white rectangle, indexed in the respective x3M, M1 and M2 settings after the analysis presented in (d-g). The rutile-like spots indicated by blue circles correspond to a coincident site lattice (CSL) defined by overlapping the x3M, M1 and M2 domain orientations coexisting in the tweed. (c) Filtered enlarged view of (a). Arrows indicate contrast modulations tripling (x3M) and doubling (M1) the rutile periodicity. The yellow rectangle in the inset highlights the preservation of the body-centered tetragonal arrangement of atomic columns underlying the contrast modulations. (d-g) Calculated electron diffraction patterns of M2 (d,e) and M1 (f,g) domain orientations contributing to the FFT shown in (b). Additional $(0\ 1\ 0)_R$,$(0\ 0\ 1)_R$, $(0\ ½\ ½)_R$ spots are indicated in red color. The orientation of $m_y$ in (d-f) coincides with the orientation of the mirror plane in the 2/*m* point group, which makes these orientations hardly discernible from equivalent rutile ones.

# Supporting Information

## Metallic Diluted Dimerization in $VO_2$ Tweeds

*Felip Sandiumenge*, Laura Rodríguez, Miguel Pruneda, César Magén, José Santiso, Gustau Catalan*

### S1: Thickness of the films

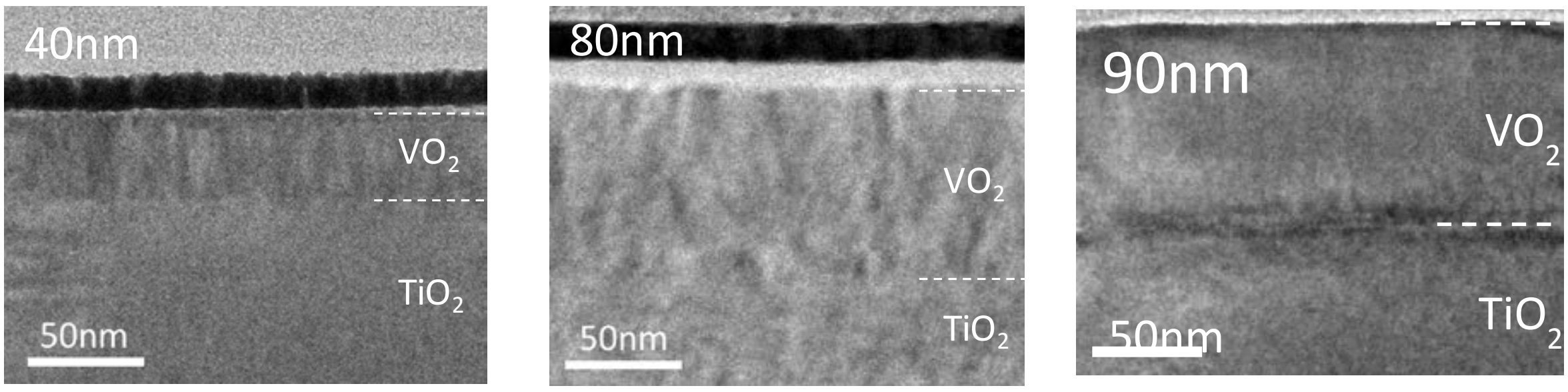

**Figure S1.** Low magnification TEM images of the three films investigated in this work showing their respective thicknesses.

**S2: Transport properties of metallic $VO_2$ domains**

a

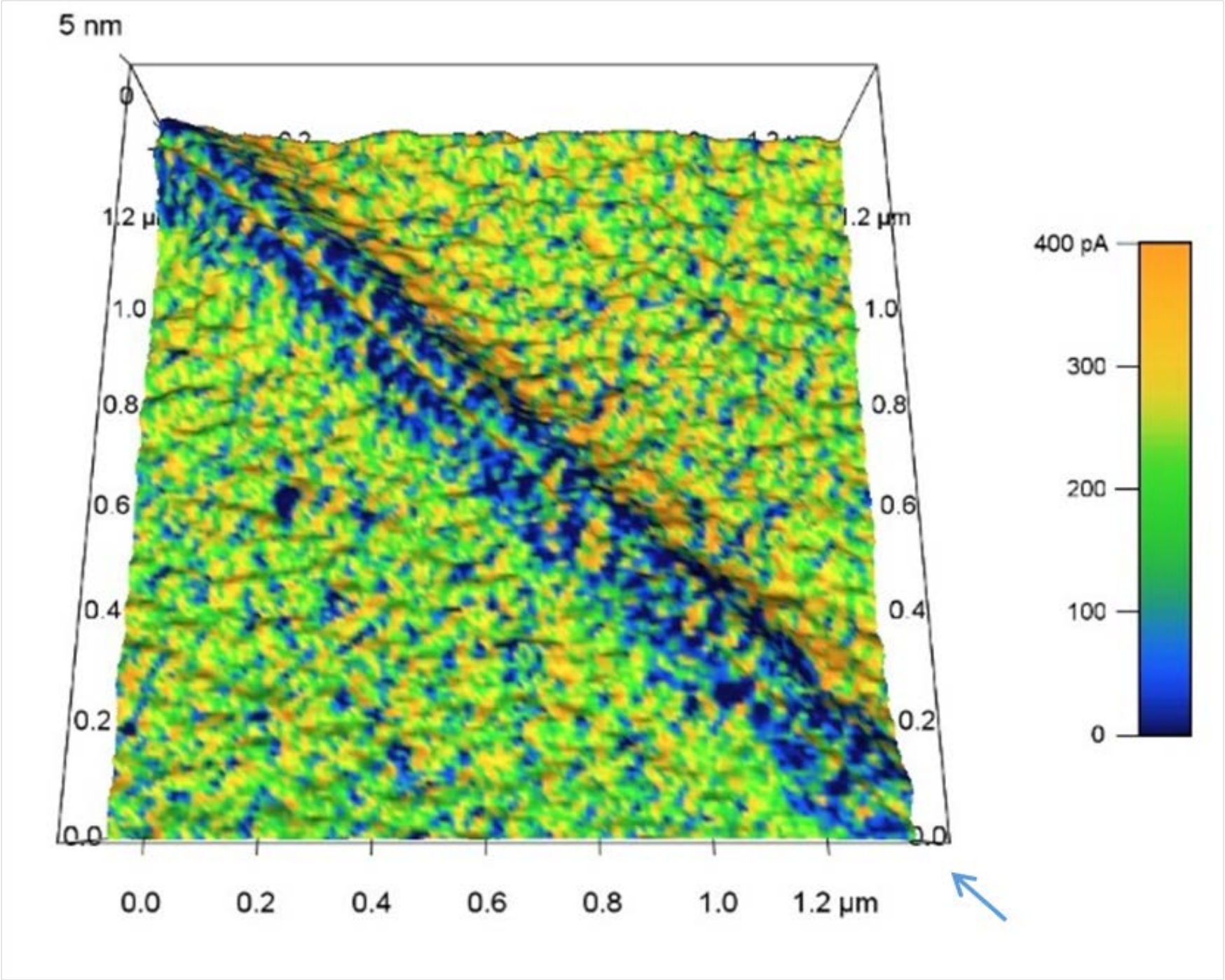

b

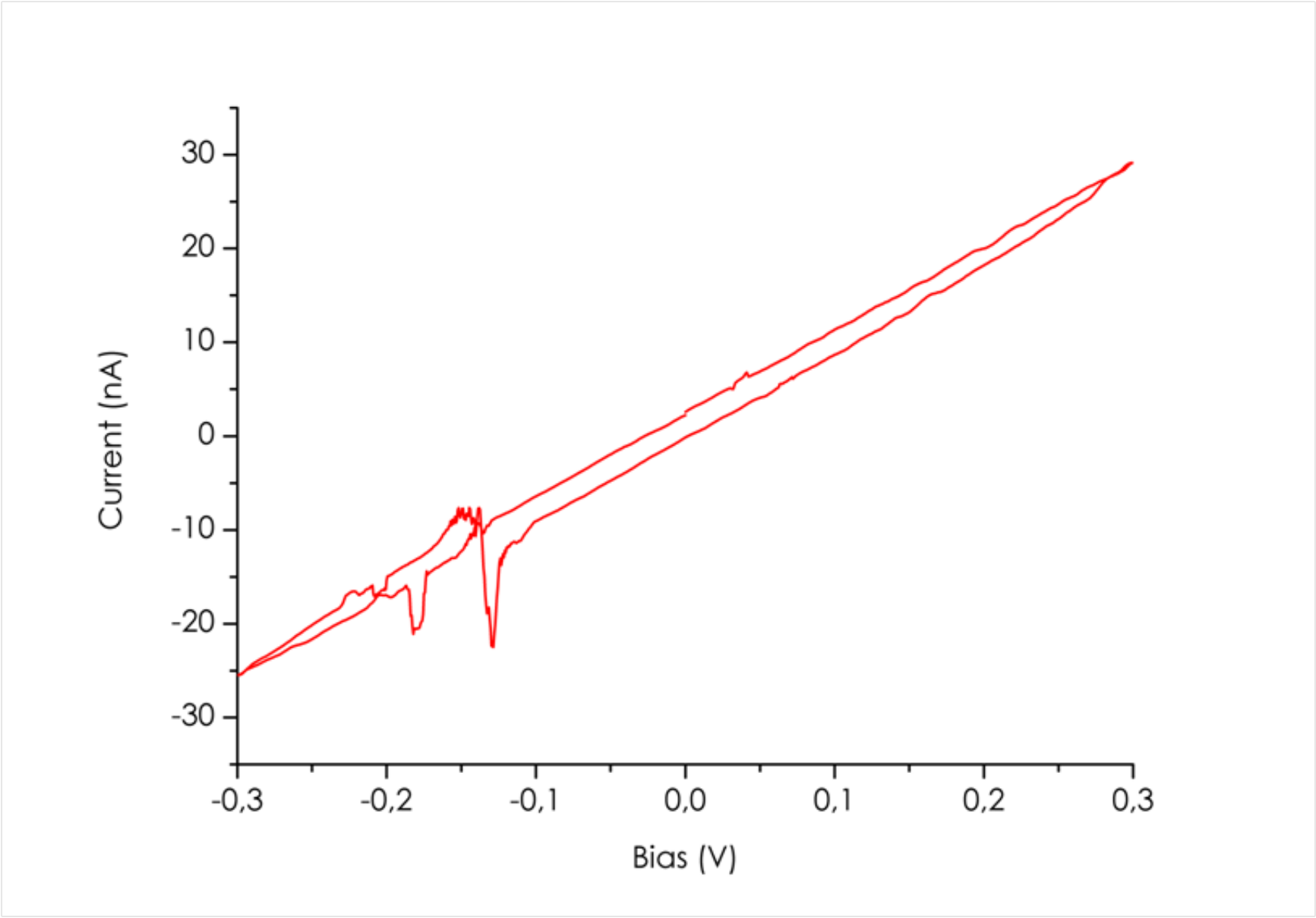

**Figure S2.** Metallic behavior of strained tweed areas between cracks. (a) Topographic conducting-AFM current map obtained from a fully strained ~36nm thick film over an area including a crack. The scan is performed in a perpendicular current configuration (sample voltage of 77 mV) using a conducting Nb-doped $TiO_2$ substrate, and shows the amount of current passing through the thickness of the film as detected by the grounded AFM tip. The arrow indicates the crack direction. The rms roughness on the regions at both sides of the crack is ~0.3nm. The crack induces a slight elevation of the film surface along its path. The dark blue color decorating both sides of the crack-line corresponds to the insulating monoclinic M1 phase bounding metallic tile-like x3M tweed regions (see Figure 1b, main text). The current signal (orange) observed along  segments of the crack gap, dividing the insulating region into two longitudinal halves, is collected from the underlying metallic substrate. (b) I(V) curve obtained from the metallic region at 15ºC.  Note that the curve exhibits a lineal behavior indicating ohmic (metallic) conduction even below room temperature.

**S3: Structure of the 90 nm -thick film**

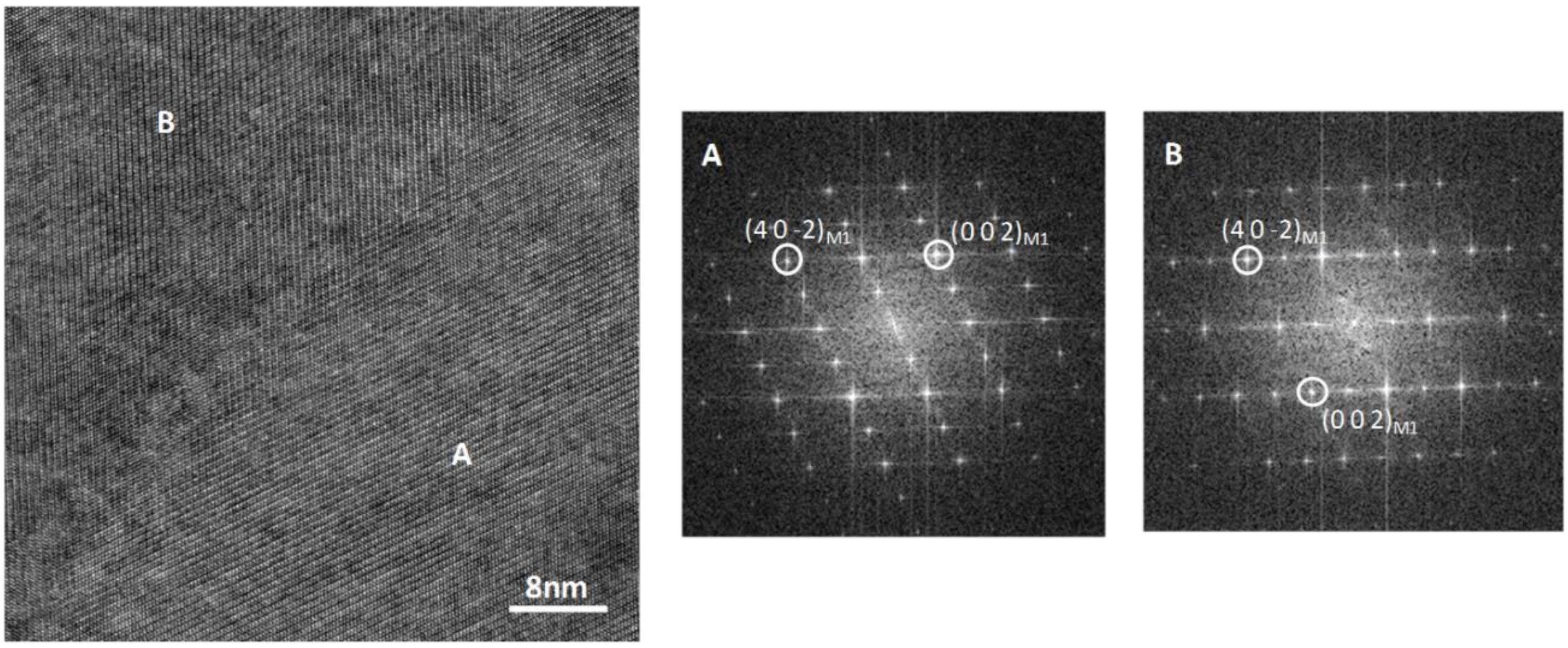

**Figure S3.** HRTEM image of the 90 nm -thick $VO_2$ film viewed along the $[0\ 1\ 0]_{M1}$ ($[1\ 0\ 0]_R$) zone axis (left panel). XRD data indicates that this is a fully relaxed M1 single phase film (Figures 1a,e in main text). The HRTEM image shows two domains, A and B, with high structural perfection. FFTs obtained from both domains, shown in the central and right panels, indicate that they are symmetrically related by the (0 1 0)R rutile mirror plane, $m_y$, lost in the rutile→M1 transition. The absence of the multicomponent tweed textures and the x3M phase observed in the strained 40 nm and 80 nm -thick films (Figures 2a and 4a of the main text, respectively) demonstrate their role in the accommodation of misfit strains up to large film thicknesses .

**S4: Crystal structure data of the x3M phase**

The unit cell parameters derived by free relaxation without imposing any symmetry constrain on the DFT calculation are $a_{x3M}$ = 8.40Å, $b_{x3M}$ = 4.71Å, $c_{x3M}$ = 5.49Å, $\beta_{x3M} \approx$ 120.7°.

| | *x* | *y* | *z* |
|---|---|---|---|
| **V1** | 0.371 | 0.902 | 0.072 |
| **V2** | 0.372 | 0.426 | 0.573 |
| **V3** | 0.682 | 0.931 | 0.113 |
| **V4** | 0.681 | 0.398 | 0.681 |
| **V5** | 0.024 | 0.903 | 0.068 |
| **V6** | 0.026 | 0.422 | 0.570 |
| **01** | 0.257 | 0.708 | 0.283 |
| **O2** | 0.455 | 0.130 | 0.873 |
| **O3** | 0.258 | 0.620 | 0.784 |
| **O4** | 0.122 | 0.220 | 0.394 |
| **05** | 0.595 | 0.698 | 0.301 |
| **O6** | 0.791 | 0.114 | 0.890 |
| **O7** | 0.595 | 0.630 | 0.801 |
| **O8** | 0.455 | 0.199 | 0.373 |
| **O9** | 0.92 | 0.722 | 0.280 |
| **O10** | 0.122 | 0.108 | 0.893 |
| **O11** | 0.925 | 0.606 | 0.780 |
| **O12** | 0.791 | 0.214 | 0.389 |

**Table S4.** Fractional atomic coordinates of the x3M crystal structure derived from DFT.

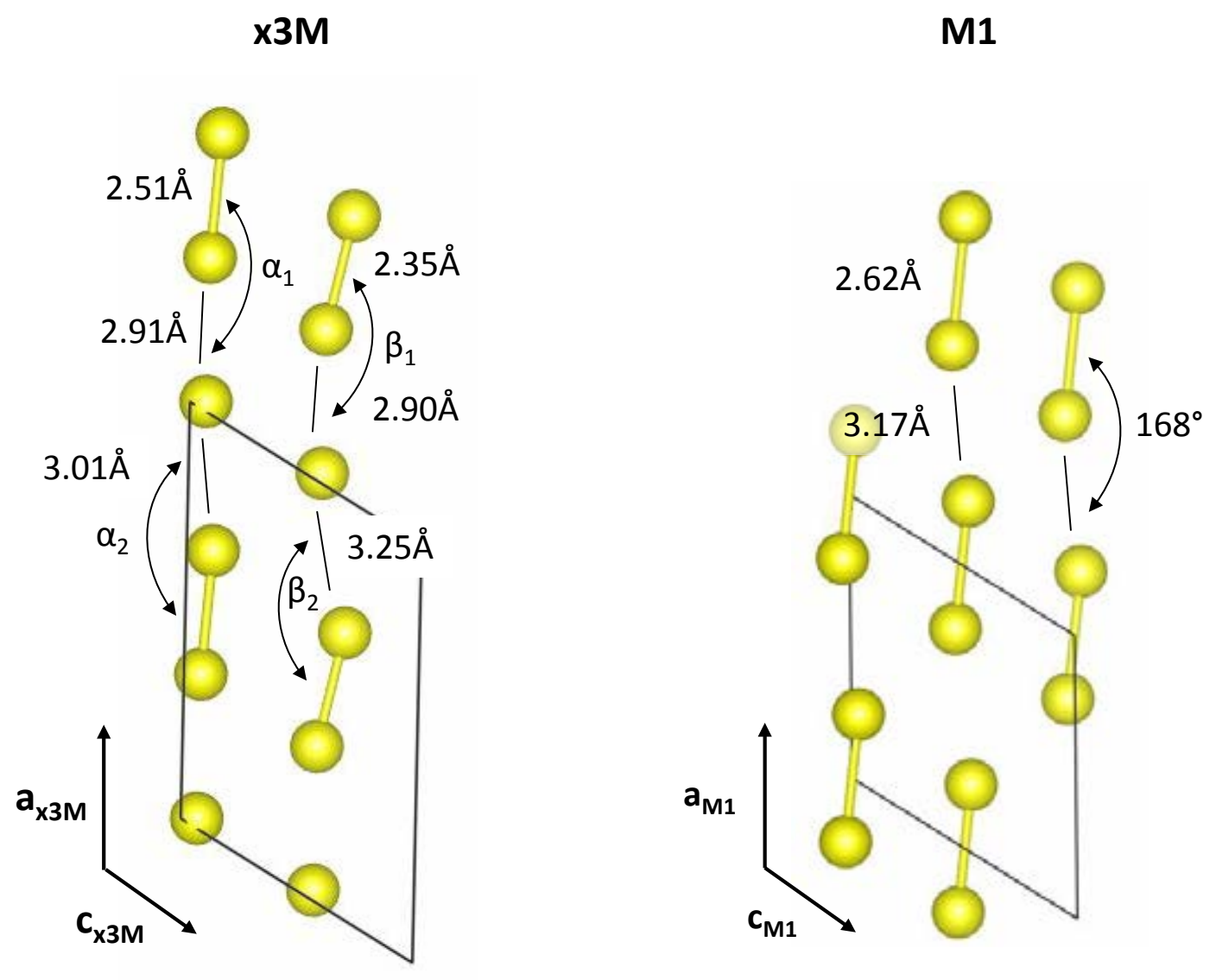

**Figure S4.** The two V-chain sublattices of the x3M structure (left panel) are unequally distorted both keeping a high short to large V-V distance ratio and V-V dimer tilts ($\alpha_1$ = 175°, $\alpha_2$ = 170°, $\beta_1$ = 167°, $\beta_2$ = 158°) as in the M1 structure (left panel, experimental data from ref.[1]. DFT data, see main text and references therein, yield slightly different values). V atoms are shown as yellow circles. V-V dimers are connected by yellow bars.

**S5: Lattice relations between the rutile and the x3M, M1 and M2 phases**

According to our DFT calculations, the relation between the x3M and rutile basis vectors is given by,

$$\mathbf{a}_{x3M} = 3\mathbf{c}_R,\ \mathbf{b}_{x3M} = \mathbf{a}_R,\ \ \mathbf{c}_{x3M} = \mathbf{b}_R - \mathbf{c}_R,$$

from which the relation between equivalent zone-axes, [u v w], is given by:

$$u_{x3M} = ⅓v_R + ⅓w_R,\ v_{x3M} = u_R,\ w_{x3M} = v_R \qquad (1).$$

According to equation (1), the x3M zone axes along the equivalent Rutile $[1\ 0\ 0]_R$ and $[0\ 1\ 0]_R$ ones are $[0\ 1\ 0]_{x3M}$ and $[1\ 0\ 3]_{x3M}$, respectively.

For the M1 and M2 phases, these relations are:[2]

$$\mathbf{a}_{M1} = 2\mathbf{c}_R,\ \mathbf{b}_{M1} = \mathbf{a}_R,\ \ \mathbf{c}_{M1} = \mathbf{b}_R - \mathbf{c}_R$$

$$u_{M1} = ½v_R + ½w_R,\ v_{M1} = u_R,\ w_{M1} = v_R \qquad (2),$$

and,

$$\mathbf{a}_{M2} = 2\mathbf{a}_R,\ \mathbf{b}_{M2} = 2\mathbf{c}_R,\ \ \mathbf{c}_{M2} = -\mathbf{b}_R$$

$$u_{M2} = ½u_R,\ v_{M2} = -w_R,\ w_{M2} = ½v_R \qquad (3),$$

respectively. According to equations (2) and (3), the equivalent Rutile $[1\ 0\ 0]_R$ and $[0\ 1\ 0]_R$ zones are, for M1: $[0\ 1\ 0]_{M1}$, $[1\ 0\ 2]_{M1}$, and for M2: $[1\ 0\ 0]_{M2}$, $[0\ 0\ 1]_{M2}$.

These unit cell relations are plotted in Figure S3 along with their 2/*m* point symmetries projected along the respective zone-axes.

**a**

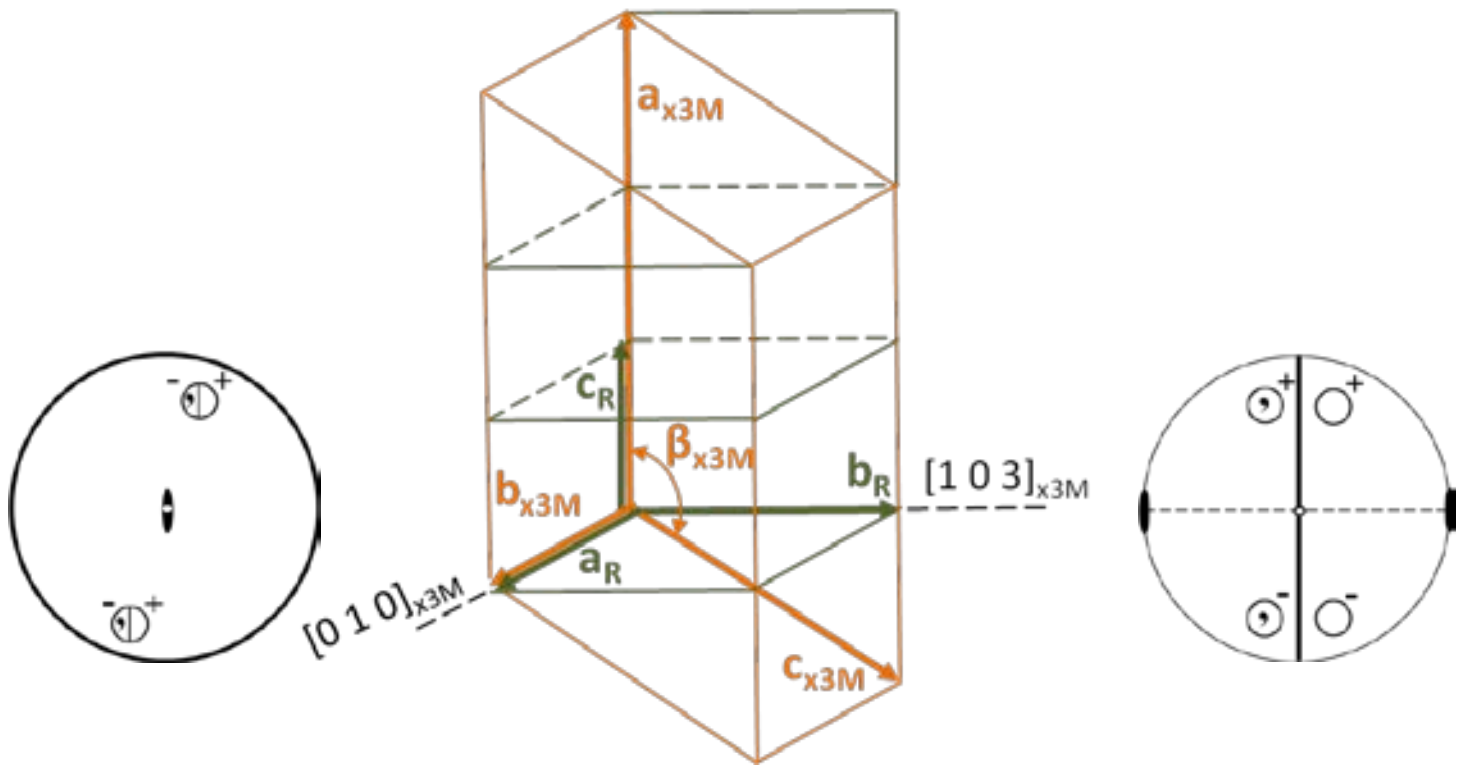

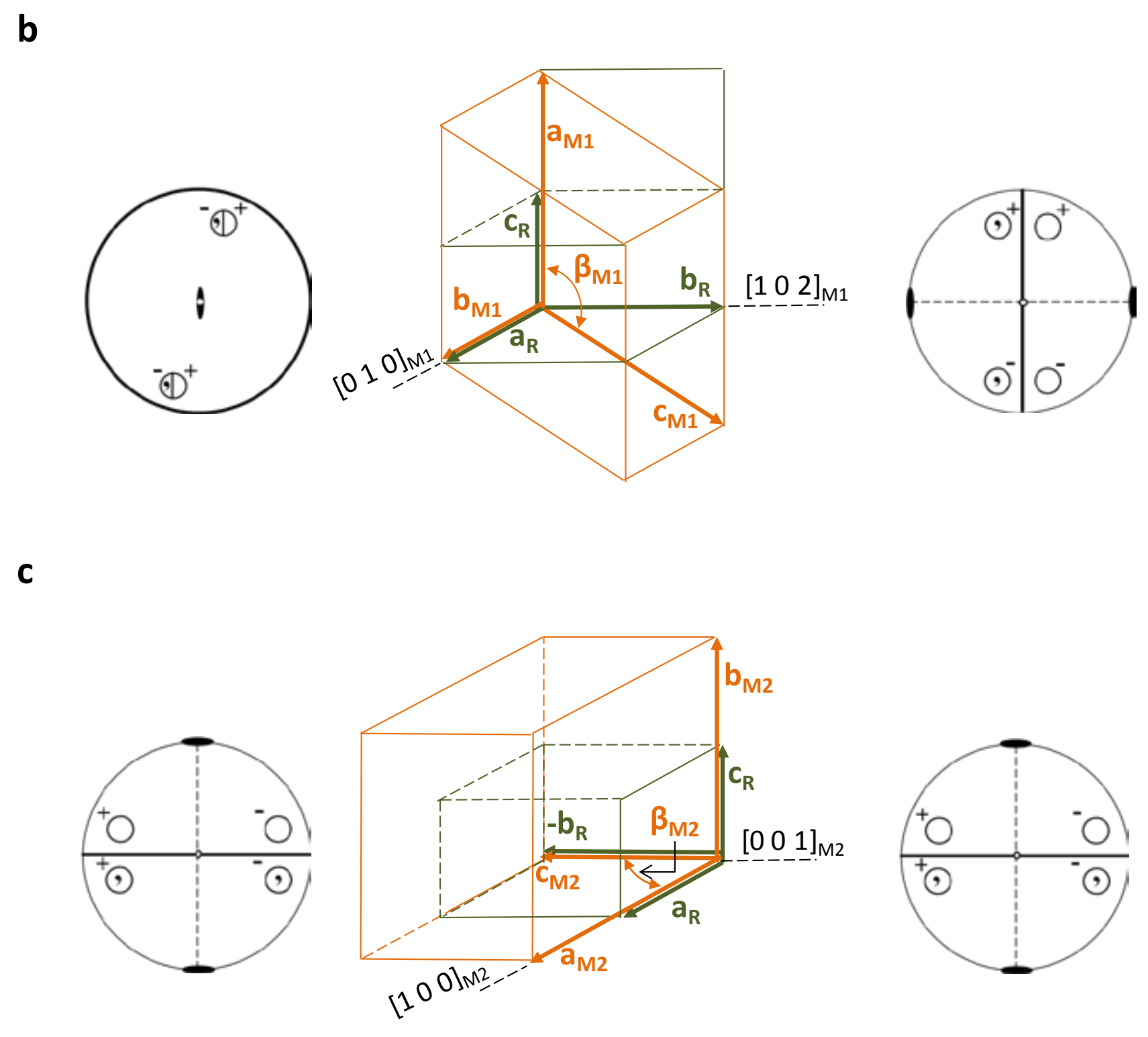

**Figure S5.** Unit cell relations between the rutile and the x3M, M1 and M2 structures, and respective zone axis symmetries. (a) x3M: the rutile [1 0 0]$_R$ zone axis is parallel to the monoclinic unique axis (two-fold axis in the 2/*m* point grup); therefore this projection lacks the mirror symmetry. Conversely, the [0 1 0]$_R$ zone axis includes this mirror symmetry. (b) M1: the unit cell relations and the zone axis symmetries are similar to the x3M case shown in (a). (c) M2: in this case the monoclinic unique axis is aligned with $c_R$ and both, the [1 0 0]$_{M2}$ and [0 0 1]$_{M2}$ zones axes exhibit mirror symmetry. (Symbolism used in point group stereographic projections: mirror planes are shown as bold lines, the two-fold axis as an ellipse and the symmetry center as white dot. A comma (,) inside the circles indicates a right to left-handed transformation by mirror reflection, - and + signs indicate that the motif is below or above the projection plane, and a vertical bar indicates that two (-,+) circles are superposed along the viewing direction).

**S6: Domain orientations of the x3M, M1 and M2 phases**

According to our DFT calculations, the x3M phase has a 2/*m* monoclinic point symmetry, such as the M1 and M2 polymorphs. From group theory, the number of symmetry-degenerated orientations is equal to the order of the prototypic point group (4/*mmm*, order 16) divided by the order of the ferroelastic point group (2/*m*, order 4).[3] Therefore, all the three x3M, M1 and M2 phases exhibit four domain orientations. In Figures S3a,b it can be seen that in the x3M and M1 structures the unique axis is perpendicular to the rutile unique axis ($\parallel \mathbf{c_R}$), while in the M2 structure the unique axis is parallel to $\mathbf{c_R}$ (Figure S3c). The symmetry of the different domain projections determines to a large extent their contrast in HRTEM images. Projections along the monoclinic unique axis, lacking the mirror symmetry, clearly discern the long and short V-V distances (Figures 4a,c) and exhibit characteristic {0 1 1}R contrast fringes in the HRTEM images, tripling or doubling the periodicity in the x3M or M1 phases, respectively, as observed in Figure 4c (main text). The combination of these domain orientations with their respective $m_y$-mirror mates ($m_y$ is referred to the rutile structure) produces the composite EDPs shown in Figures S4b,d. The most intense spots of these composite patterns reproduce the tetragonal rutile motif highlighted in the FFT shown in Figure 4b (main text), and indicate the preservation of the rutile template in the tweeds. All other projections, exhibiting mirror symmetry, exhibit only small deviations from the tetragonal symmetry. In the case of the M2 phase, owing to the coincidence of the monoclinic and rutile unique axes, this situation holds for the two equivalent zone axes (see Figures S4c and S4e). HRTEM images of these domain orientations are similar and retain the main contrast features of the prototypic rutile structure.

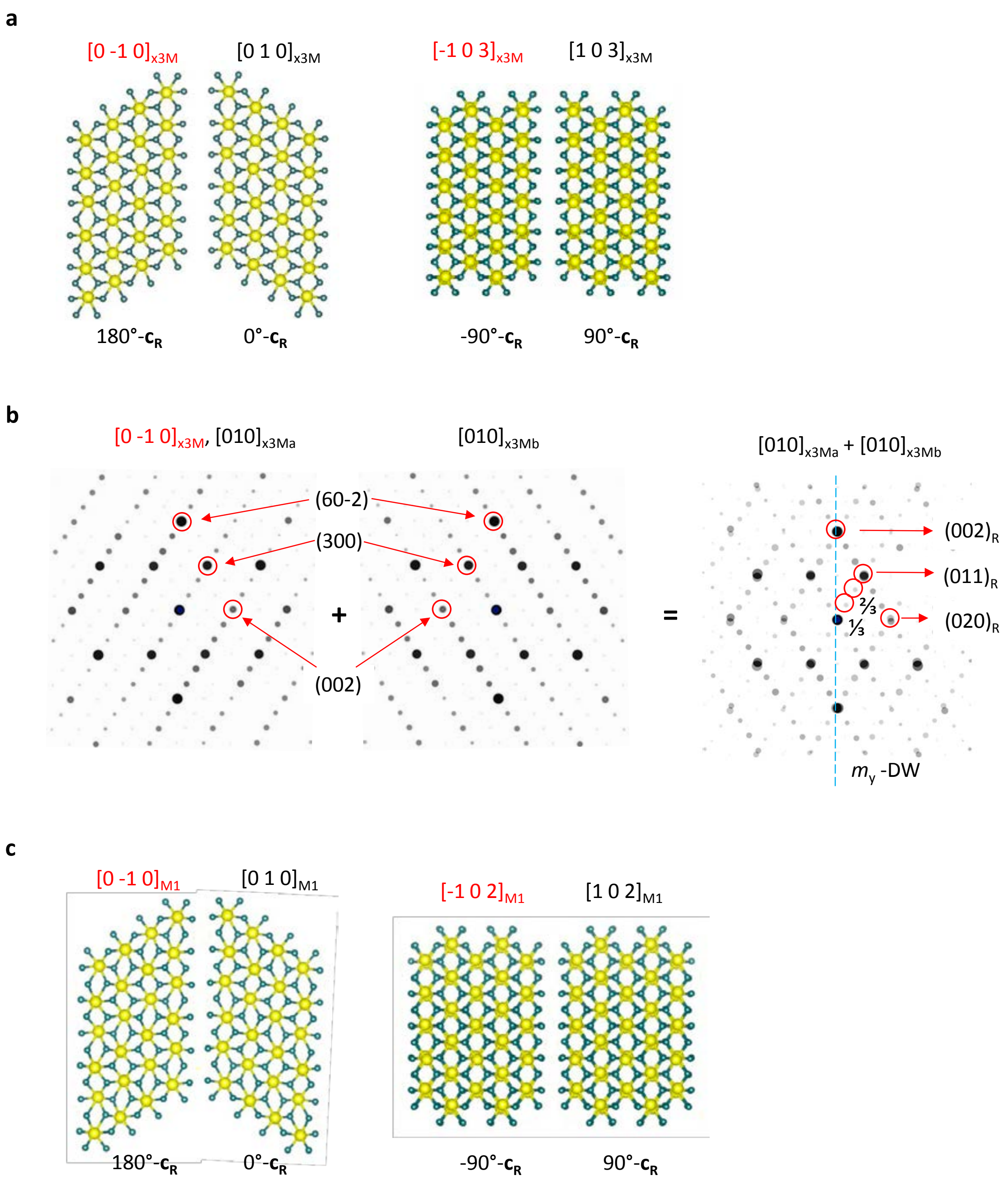

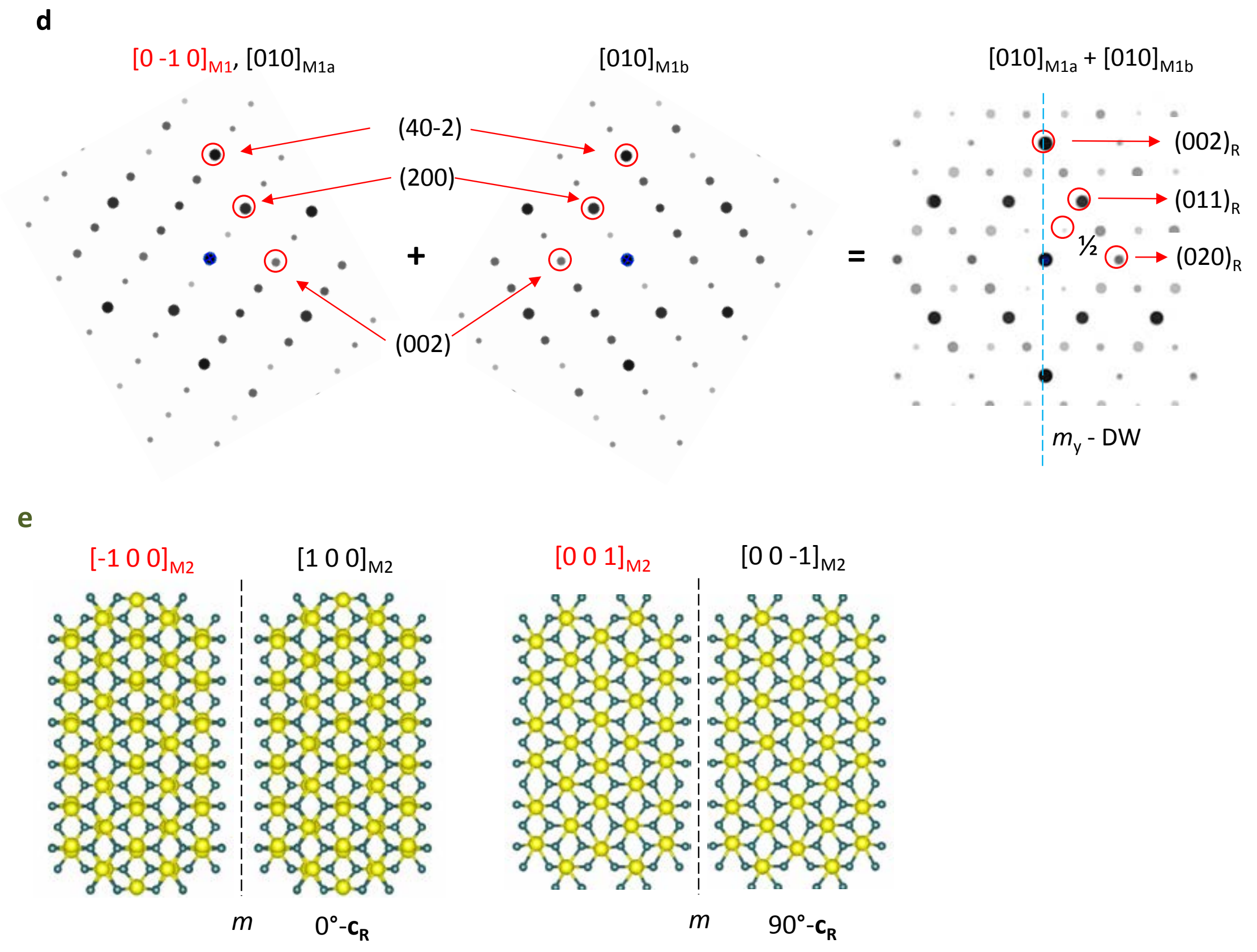

**Figure S6.** (a) Domain orientations of the x3M phase. Since the unique monoclinic axis is perpendicular to unique tetragonal axis (Figure S4a), the four domains in the x3M phase are generated by 90° rotations around the $\mathbf{c}_R$-axis. The left panel shows the $[1\ 0\ 0]_R$ projections of 0° and 180° domains, corresponding to the $[0\ 1\ 0]_{x3M}$ and $[0\ {-1}\ 0]_{x3M}$ orientations, respectively. These domains are referred to as "a" and "b" in the main text, and are related by the rutile $m_y$ mirror plane lost in $4/mmm \rightarrow 2/m$ transition. As observed in the structure projections, atom columns are aligned with the zone axis and therefore the intra (S) and inter (L) V-V distances are clearly discernible, giving rise to the $\{0\ 1\ 1\}_R$ x3 fringe contrast observed in the HRTEM images. The right panel shows projections of the 90° and -90° x3M domains, corresponding to the $[1\ 0\ 3]_{x3M}$ and $[-1\ 0\ 3]_{x3M}$ zone axes, respectively. This projections have a higher symmetry than the 0° and 180° ones (see Figure S4a) and the tetragonal rutile motif is undistorted. This symmetry constrain makes these domains hardly distinguishable from the rutile structure in the HRTEM images (see Figure 3d, main text). (b) Composite $[0\ 1\ 0]_{x3Ma}$ + $[0\ 1\ 0]_{x3Mb}$ EDP shown in Figure 3f of main text, obtained as the sum of those corresponding

to the 0° and 180°x3M domains. In the left panels, main spots are indexed in the x3M lattice. In the right panel, main spots are indexed in the rutile lattice. $(0\ \frac{1}{3}\ \frac{1}{3})_R$ and $(0\ \frac{2}{3}\ \frac{2}{3})_R$ spots are indicated by ⅓ and ⅔, respectively. It can be seen that *the composite pattern defines a coincidence site lattice which reproduces the basic Rutile diffraction motif.* The vertical dashed line traces the rutile $m_y$ mirror plane which defines the $m_y$ domain wall ($m_y$-DW) orientation between both domains in real-space.[4] (c) Domain orientations of the M1 phase. Since the unit cell holds the same orientation relation with the rutile lattice as the x3M phase, the same considerations above (a) apply in this case. The fringe contrast observed in the $[0\ 1\ 0]_R$ HRTEM images doubles, instead of tripling, the $\{0\ 1\ 1\}_R$ periodicity. (d) Composite $[0\ 1\ 0]_{x3Ma}$ + $[0\ 1\ 0]_{x3Mb}$ EDP shown in Figure 4g of main text, obtained as the sum of those corresponding to the 0° and 180°M1 domains. In the left panels, main spots are indexed in the M1 lattice. In the right panel, main spots are indexed in the rutile lattice. The $(0\ \frac{1}{2}\ \frac{1}{2})_R$ spot is indicated by ½. It can be seen that, as for the x3M phase, the composite pattern defines a coincidence site lattice which reproduces the basic rutile diffraction motif highlighted in Figure 4b of main text. The vertical dashed line traces the rutile $m_y$ mirror plane which defines the $m_y$ domain wall ($m_y$-DW) orientation between both domains in real-space. (e) In contrast with the x3M and M1 phases, in this case the monoclinic and tetragonal unique axes are parallel (Figure S4c). Two out of the four domain orientations are obtained by 90° rotations around the $\mathbf{c_R}$-axis, and the other two orientations are then generated by mirror symmetry. The doubling of the prototype tetragonal motif can be clearly seen in all the projections.